\documentclass[conference]{IEEEtran}
\IEEEoverridecommandlockouts
\usepackage{cite}
\usepackage{amsmath,amssymb,amsfonts}
\usepackage{algorithmic}
\usepackage{graphicx}
\usepackage{textcomp}
\usepackage{hyperref}
\usepackage[table]{xcolor}
\usepackage{booktabs}
\usepackage[most]{tcolorbox}
\usepackage{float}
\usepackage{tikz}
\usetikzlibrary{positioning, fit, backgrounds, calc}

\def\BibTeX{{\rm B\kern-.05em{\sc i\kern-.025em b}\kern-.08em
    T\kern-.1667em\lower.7ex\hbox{E}\kern-.125emX}}
    
\begin{document}

\title{Helpful or Harmful? Evaluating LLM-Assisted Vulnerability Patching via a Human Study}
% \title{AI Assistance in Vulnerability Patching\\}

\author{
\IEEEauthorblockN{
Giulian Biolo\textsuperscript{1}, 
Michael Tezza\textsuperscript{1}, 
Yuanjun Gong\textsuperscript{1}, 
Fabio Massacci\textsuperscript{1,2}
}
\IEEEauthorblockA{
\textsuperscript{1}{University of Trento, IT} \\
\textsuperscript{2}{Vrije Universiteit Amsterdam, NL} \\
\{giulian.biolo, michael.tezza\}@studenti.unitn.it, yuanjun.gong@unitn.it, fabio.massacci@ieee.org
}
}

% \author{
% \IEEEauthorblockN{Giulian Biolo}
% \IEEEauthorblockA{\textit{University of Trento, IT} \\
% giulian.biolo@studenti.unitn.it}
% \and
% \IEEEauthorblockN{Michael Tezza}
% \IEEEauthorblockA{\textit{University of Trento, IT} \\
% michael.tezza@studenti.unitn.it}
% \and
% \IEEEauthorblockN{Yuanjun Gong}
% \IEEEauthorblockA{\textit{University of Trento, IT} \\
% yuanjun.gong@unitn.it}
% \and
% \IEEEauthorblockN{Fabio Massacci}
% \IEEEauthorblockA{\textit{University of Trento, IT} \\
% \textit{Vrije Universiteit Amsterdam, NL}\\
% fabio.massacci@ieee.org}
% }

\maketitle

\begin{abstract}
\noindent \textbf{[Background:]} Software vulnerability remediation is a cognitively demanding task that requires specialized security expertise often lacking in general developers. In the meantime, Large Language Models (LLMs) assisted tools show potential in vulnerability detection, location, and repair tasks. 
\textbf{[Hypothesis:]} While LLM-assistance is hypothesized to accelerate patching, it also risks introducing \emph{hallucinations} or insecure code, leading to a higher likelihood of generating superficial repairs that bypass the standard functionality checks but fail the security validation.
\textbf{[Objective:]} We aim to present an empirical experiment, unveiling the capability of LLM-assisted vulnerability patching compared to manual debugging on human participants in real-world scenarios. 
\textbf{[Method:]} We plan to conduct a controlled experiment using a Balanced Crossover design. For that, we have developed a WebApp for code execution and integrated hidden \emph{Ghost Tests} to verify patch integrity beyond visible functional requirements. The experiment involves training and evaluation scenarios. The remediation speed, remediation efficacy for both standard functionality tests and security tests, and participant perception will be evaluated. 
\textbf{[Pilot Study:]} A pilot experiment with a small sample of participants has been conducted, providing insights for the following study.

\end{abstract}


\section{Introduction}
Software vulnerabilities represent a persistent threat to information systems, where the speed of remediation is often critical to preventing exploitation. Yet, manual vulnerability remediation remains a cognitively intensive task requiring deep codebase knowledge and specialized security principles expertise, often lacking in general developers working under time pressure. The advent of LLMs and LLM-assisted  coding tools has increasingly fueled the hypothesis that these technologies can significantly accelerate the remediation process \cite{wu2023effective,pearce2023examining,wang2025vulnrepaireval,bhatt2023cyberseceval}.

Relying on LLMs for secure code generation, however, is challenging. Recent studies and experimental results suggest that while an LLM can generate code patches rapidly, it may also introduce subtle vulnerabilities or ``hallucinations'' if not rigorously verified \cite{sajadi2025secure,baker2025monitoring,fu2025security,bhatt2023cyberseceval}. In the realm of human-subject studies, a mixed yet concerning picture is presented regarding the security of LLM-assisted code generation. LLM assistants  are charged with the tendency of producing less secure code, and yet paradoxically feeling more confident in their security, causing potentially insecure dependencies from human users~\cite{perry2023users,serafini2025exploring,klemmer2024using}.
On the other hand, the perceptions of the security of LLM-generated code also vary depending on task complexity\cite{sandoval2023lost}, vulnerability type\cite{asare2024user}, expected fixing methods~\cite{steenhoek2025closing}, and developer over-reliance dynamics that persist even under explicit warnings or recommended interventions~\cite{perry2023users,klemmer2024using,serafini2025exploring}.

In our study, we aim to investigate the functionality and security of LLM-generated patches for vulnerabilities. We argue that the risk of \emph{fake fixes} illusion is at play, which means that LLMs generate code that successfully passes basic functional tests, thereby satisfying the developer, while leaving the underlying security flaw entirely intact.

\begin{tcolorbox}[colback=gray!3, colframe=black!30, boxrule=0.4pt, arc=1mm, boxsep=0.5pt]
\noindent \textbf{Central Hypothesis.} LLM assistance reduces remediation time but leads to lower security effectiveness, while being perceived as helpful by users.
\end{tcolorbox}
\vspace{0.5em}

Throughout, the unit of analysis is the human--LLM interaction: we measure how participants direct, judge, and adapt to LLM assistance, not the accuracy or stability of any individual model. We therefore raise three research problems:
\begin{itemize}
    \item \textbf{RQ1.} How does LLM assistance affect the time required for vulnerability diagnosis and repair? (Speed)
    \item \textbf{RQ2.} How does the use of an LLM affect the quality of the patch when subjected to security tests? (Efficacy)
    \item \textbf{RQ3.} How do developers perceive the helpfulness of LLM tools when solving complex security tasks? (Perception)
\end{itemize}

We propose a controlled empirical study protocol to investigate the true impact of LLM assistants on the security patching process. 
To expose these \emph{fake fixes} and test our hypothesis, we plan to measure the patch's resilience against exploits that the developer (and the LLM) are not explicitly optimizing for. Therefore, we systematically introduce \emph{Ghost Tests}, security-specific test cases that evaluate whether a patch genuinely resolves the vulnerability or merely circumvents visible functional requirements.

Operationally, the experiment runs on a custom-developed WebApp providing code execution and telemetry, combined with LimeSurvey for demographic profiling. We compare a manual approach (relying on standard documentation) against an LLM-assisted one (using a general-purpose LLM coding assistant), to determine whether LLMs improve remediation speed without compromising the security of the code, and how developers perceive their usefulness.

\section{Terminology and Background}
\label{sec:background}
We frame \emph{vulnerability remediation} as the task of transforming a code artifact known to contain an exploitable flaw into a patched version that preserves the intended functional behavior while eliminating the underlying weakness. In an LLM-assisted setting, the developer interacts with a general-purpose LLM coding assistant that can diagnose, explain, or generate candidate patches; the developer retains responsibility for reviewing and integrating the model's output.

\paragraph{Scenario}
A \emph{scenario} is a self-contained vulnerability-repair task. It bundles a vulnerable Python source artifact under repair, a participant-facing description that scaffolds the discovery and triage process, a visible suite combining functional regression tests with security tests that target a specific exploit, and a hidden security-oriented suite that probes orthogonal exploit-relevant behavior, unseen by the participant during the task.

\paragraph{Fake Fixes and Ghost Tests}
A central concern of this study is the \emph{fake fix}: a patch that satisfies the visible test suite without removing the root cause of the vulnerability. Fake fixes arise both from human shortcuts (e.g., hardcoding the expected output to satisfy assertions) and from LLM hallucinations that pattern-match to a syntactically plausible fix without grounding it in security semantics~\cite{agarwal2025codemirage,aleithan2024swebenchplus}, an instance of the broader \emph{patch overfitting} phenomenon long observed in automated program repair~\cite{petke2024patch}. \emph{Ghost Tests} are hidden, security-specific test cases, executed once at submission, that probe exploit-relevant behavior beyond functional correctness, in line with the exploit-based verification of recent benchmarks~\cite{wang2025vulnrepaireval}.

\paragraph{Balanced crossover design}
A balanced crossover design, inspired by Taguchi orthogonal arrays~\cite{mitra2011taguchi,massacci2024addressing}, exposes each participant to both manual and LLM-assisted treatments across a sequence of scenarios, with assignment ordered so that every scenario is solved by some participants under each condition. This isolates the effect of LLM assistance  from inter-participant skill variance and neutralizes carry-over and learning effects without requiring the impractically large sample size of a fully between-subjects design. The concrete assignment matrix is presented in Section~\ref{sec:methodology}.

\section{Methodology}
\label{sec:methodology}

We designed a controlled experiment orchestrated through a custom-developed WebApp that handles code execution and telemetry, together with LimeSurvey for demographic profiling and qualitative data collection.

\subsection{Methodology Overview}
Our study is structured into five phases:

\emph{\textbf{Scenario Composition:}} We first construct standardized experimental scenarios from real-world projects, each a self-contained task bundling the four artifacts of Section~\ref{sec:background}. The description documents the intended functional behavior, including an execution flow-chart; the type and location of the vulnerability are not disclosed, as diagnosis is part of the task. Six scenarios were selected and implemented in the pilot study.

\emph{\textbf{Experimental Platform Development:}} The visible suites were manually authored to high branch coverage of the core logic. The vulnerable source, the visible suite, and the LLM-assistance interface are provisioned through the WebApp to standardize the environment. Participants may run the visible tests without limit, each run logged by telemetry; the hidden Ghost Tests run once at final submission, with no feedback shown, to expose potential \emph{fake fixes}~\cite{agarwal2025codemirage}.

\emph{\textbf{Participant Assignment:}} Participants are assigned to one of two groups according to the Taguchi-inspired balanced crossover design (Section~\ref{sec:background}); with six scenarios, the assignment is shown in Table~\ref{tab:pipeline}. It guarantees that each participant solves two evaluation scenarios manually and two with LLM assistance, and that each evaluation scenario receives both treatments across the pool.

\begin{figure}[t]
    \centering
    \vspace{0.5em}
    \resizebox{0.9\columnwidth}{!}{%
    \begin{tikzpicture}[
        block/.style={
            rectangle, draw, rounded corners=2mm,
            minimum height=0.9cm, minimum width=2.1cm,
            align=center, font=\small
        },
        container/.style={
            rectangle, draw, rounded corners=3mm, inner sep=2mm
        }
    ]

    % left
    \node[block] (initial) at (0,0) {Initial\\Survey};

    % middle-left group
    \coordinate (testc) at (3.6,0);
    \node[block] (test1) at ($(testc)+(0,0.55)$) {Training Scenario S1};
    \node[block] (test2) at ($(testc)+(0,-0.55)$) {Training Scenario S2};
    \node[container, fit=(test1) (test2)] (test_container) {};

    % middle-right group
    \coordinate (scenc) at (7.1,0);
    \node[block] (scen1) at ($(scenc)+(0,1.65)$) {Scenario S3};
    \node[block] (scen2) at ($(scenc)+(0,0.55)$) {Scenario S4};
    \node[block] (scen3) at ($(scenc)+(0,-0.55)$) {Scenario S5};
    \node[block] (scen4) at ($(scenc)+(0,-1.65)$) {Scenario S6};
    \node[container, fit=(scen1) (scen2) (scen3) (scen4)] (scen_container) {};

    % right
    \node[block] (final) at (10.1,0) {Final\\Survey};

    % horizontal arrows
    \draw[->, thick] (initial.east) -- (test_container.west);
    \draw[->, thick] (test_container.east) -- (scen_container.west);
    \draw[->, thick] (scen_container.east) -- (final.west);

    \end{tikzpicture}%
    }
    \vspace{0.5em}
    \caption{Experimental workflow from participant onboarding to patch evaluation and metric extraction.}
    \label{fig:method_workflow}
\end{figure}
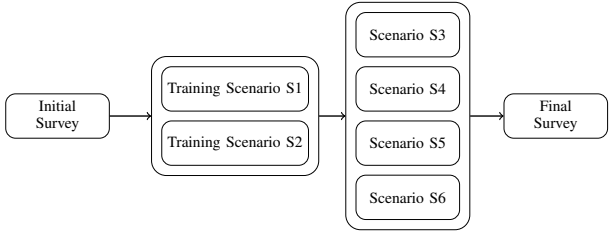

\emph{\textbf{Participant Task Procedure:}} The participant task comprises an initial demographic survey (5 mins), two training scenarios (S1--S2, 60 mins), four evaluation scenarios (S3--S6, 120 mins), and a post-experiment survey (5 mins). Each scenario is subject to a hard 30-min time box, recording a timeout if nothing is submitted; participants can inspect the vulnerable code, apply their assigned treatment (manual or LLM-assisted), submit and re-submit patched code, run the visible tests, or terminate the scenario early. Web access is permitted in both conditions; the LLM assistant is permitted only in the LLM-assisted condition. The workflow is illustrated in Figure~\ref{fig:method_workflow}.

\emph{\textbf{Data Collection and Analysis:}} A Supabase database stores the experimental variables summarized in Table~\ref{tab:variables}; the data are analyzed with the statistical methods of Section~\ref{subsec:data_analysis}.

\subsection{Hypotheses and Variables}
\label{subsec:metrics}

The descriptions of the experimental variables are summarized in Table~\ref{tab:variables}. \emph{Independent} variables (intervention type, scenario identifier) define the experimental design; \emph{dependent} variables operationalize the constructs of speed, efficacy, and perception; \emph{confounding} variables capture self-reported background and are used to control for skill in the analysis. 

\subsubsection{RQ1} How does LLM assistance affect the time required for vulnerability diagnosis and repair? (Speed)

Let $T_{\mathrm{open}}$ be the timestamp at which the participant opens a scenario and $T_{\mathrm{sub}}$ the timestamp of their final compiling submission within the time box. The remediation time for that scenario is

\begin{equation}
    T = \min\bigl(T_{\mathrm{sub}} - T_{\mathrm{open}},\; 30\text{ min}\bigr)
    \label{eq:m1}
\end{equation}

\textbf{$H_T$:} The LLM-assisted condition yields a lower remediation time $T$ than the manual condition.

\begin{table*}[tbp]
  \caption{Experiment Design (Taguchi Balanced Crossover)}
  % \vspace{-8pt}
  % \longcaption{0.65\linewidth}{The table shows the intervention assigned to each group for each task. \textbf{Manual} represents the Control condition, and \textbf{AI} represents the Treatment condition.}
  
  \label{tab:pipeline}
  \centering
  \footnotesize
  \begin{tabular}{|l|c|c|c|c|c|}
    \toprule
    \textbf{Group} & \textbf{Training (S1-S2)} & \textbf{Scenario S3} & \textbf{Scenario S4} & \textbf{Scenario S5} & \textbf{Scenario S6} \\
    \midrule
    \textbf{Group A} & \cellcolor{orange!20} Manual / LLMs  & \cellcolor{red!20} Manual & \cellcolor{green!20} LLMs & \cellcolor{red!20} Manual & \cellcolor{green!20} LLMs \\ 
    \textbf{Group B} & \cellcolor{orange!20} LLMs   / Manual & \cellcolor{green!20} LLMs & \cellcolor{red!20} Manual & \cellcolor{green!20} LLMs & \cellcolor{red!20} Manual \\
  \bottomrule
    \end{tabular} 
\end{table*}

\subsubsection{RQ2} How does the use of an LLM affect the quality of the patch when subjected to security tests? (Efficacy)

Let $F \in \{0,1\}$ denote passing the visible functional tests, and $S \in \{0,1\}$ denote passing all security tests, i.e., the visible security tests together with the hidden Ghost Tests. A patch is \emph{genuinely secure} only when $\{F=1, S=1\}$; a patch with $\{F=1, S=0\}$ is a fake fix in the sense of Section~\ref{sec:background}. A timed-out submission results in $\{F=0, S=0\}$.

\textbf{$H_E$: }The proportion of \emph{fake fixes} is higher in the LLM-assisted group.

To better understand the underlying causes, we further examine three aspects:

\textbf{$H_E^F$:} The pass rate of visible functional tests is higher under the LLM-assisted condition.

\textbf{$H_E^S$:} The success rate of resolving the core vulnerability is lower under the LLM-assisted condition.

\textbf{$H_E^E$:} The overall score $E$ ($E = F \cdot S$) is lower under the LLM-assisted condition.

\subsubsection{RQ3} How do developers perceive the helpfulness of LLM tools when solving complex security tasks? (Perception)

After each evaluation scenario, participants self-report the perceived helpfulness of their assigned treatment on the 5-point Likert scale of Table~\ref{tab:variables}; aggregate ratings address RQ3 independently of the objective outcomes.

\textbf{$H_P$:} Users report positive perceptions toward LLM-based repair assistance.

We further examine the perception of users with different levels of Python security patching self-reported experience:

\textbf{$H_P^H$:} Users with high self-reported Python security patching experience report positive perceptions toward LLM-based repair assistance.

\textbf{$H_P^L$:} Users with low self-reported Python security patching experience report positive perceptions toward LLM-based repair assistance.

\begin{table*}[tbp]
\caption{Experimental Variables Collected During the Task.}
\label{tab:variables}
\centering
\footnotesize
\begin{tabular}{lp{0.45\linewidth}ll}
\toprule
\textbf{Name} & \textbf{Description} & \textbf{Scale} & \textbf{Operationalization} \\
\midrule
% \multicolumn{4}{l}{\emph{Independent variables}} \\
\emph{\textbf{Independent variables}}  & \multicolumn{3}{l}{\emph{\textbf{Define the experimental design}}}\\
\midrule
Intervention Type & Method used to remediate the vulnerability. & Nominal & Binary (0=Manual, 1=LLM-Assisted) \\
Scenario ID & Specific vulnerability task assigned. & Nominal & Categorical (S1--S6) \\
\midrule
% \multicolumn{4}{l}{\emph{Dependent variables}} \\
\emph{\textbf{Dependent variables}}&\multicolumn{3}{l}{\emph{\textbf{Operationalize the constructs of efficiency,
efficacy, and perception}}}\\
\midrule
Submission Time & Time elapsed from task opening to final submission. & Ratio & Minutes (capped at 30) \\
Functional Status & Outcome of the visible unit-test suite. & Nominal & Binary (Pass/Fail) \\
Security Status & Outcome of the hidden Ghost-Test suite. & Nominal & Binary (Secure/Insecure) \\
Perceived Helpfulness & Self-reported utility of the assistance condition. & Ordinal & 5-point Likert scale \\
\midrule
% \multicolumn{4}{l}{\emph{Confounding variables}} \\
\emph{\textbf{Confounding variables}}& \multicolumn{3}{l}{\emph{\textbf{Capture self-reported background and are used to control for skill in the analysis.}}} \\
\midrule
Python Experience & Self-reported programming proficiency. & Ordinal & 1--5 (Novice to Expert) \\
Security Experience & Self-reported vulnerability-fixing experience. & Ordinal & 1--5 (None to Expert) \\
\bottomrule
\end{tabular}
\end{table*}

\subsection{Data Analysis}
\label{subsec:data_analysis}

We apply the Benjamini--Hochberg procedure to adjust $p$-values for multiple comparisons.

The Python security patching experience background of users is assessed based on the combination of two confounding variables, Python Experience and Security Experience.  
For the objective outcomes (RQ1, RQ2), the balanced within-subjects crossover holds baseline skill constant: each participant solves scenarios under both conditions, and the per-participant frailty term (RQ1) and paired tests (RQ2) absorb the remaining participant-level variation, so no separate skill stratification is needed.
For the perception analysis (RQ3), two secondary hypotheses ($H_P^H$, $H_P^L$) decompose $H_P$ by participant self-reported experience: we rank participants on the sum of the self-reported Python experience ($1-5$) and security experience ($1-5$), split at the median, and test whether each half reports positive perceptions.

\subsubsection{RQ1} Due to the time box, remediation time is right-censored. We evaluate $H_T$ with survival analysis: a Cox proportional-hazards model of time-to-submission with treatment as a fixed effect and a per-participant frailty term, censoring sessions at the time box. The frailty term accounts for the repeated-measures structure of the crossover; for the censored speed outcome it plays the role that the paired tests below play for the uncensored ones.

\subsubsection{RQ2} For $H_E$, we compare the proportion of \emph{fake fixes} between the LLM-assisted and manual conditions using a one-sided Fisher's exact test. As for the three related hypotheses, for $H_E^F$ and $H_E^S$, we compare proportions between conditions using one-sided Fisher's exact tests. Aligning with the paired design, we evaluate $E$ using a one-sided Wilcoxon signed-rank test on each participant's mean score per condition.

\subsubsection{RQ3} For $H_P$, we compare users' 1--5 Likert-scale perception ratings against the neutral midpoint of 3 using a one-sided one-sample Wilcoxon signed-rank test.
For $H_P^H$ and $H_P^L$, the median split keeps the subgroups balanced by construction. Since each test runs on about half the samples, these two tests remain secondary to the primary confirmatory perception claim $H_P$.

\section{Execution Plan}

\subsection{Scenarios Selection Criteria}
\label{subsec:scenario_criteria}

Candidate vulnerabilities are admitted to the scenario pool only if they jointly satisfy: 

\begin{itemize}
    \item[$s_1$] \emph{Real-world representativeness.} The vulnerability class belongs to the CWE Top~25 or the current OWASP Top~10, and the starter artifact is extracted from an open-source repository rather than synthesised.
    \item[$s_2$] \emph{Single-file Python, standard-library dependencies only.} Imposed by the in-browser Pyodide runtime that backs the WebApp's code execution layer.
    \item[$s_3$] \emph{Time-box feasibility.} The scenario admits a complete manual remediation within the time box. This was verified by the authors and assessed in the pilot study.
    \item[$s_4$] \emph{Ghost-Test separability.} The vulnerability admits exploit variants distinct from those probed by the visible suite, so that a hidden suite differentiating superficial from genuine fixes can be constructed. All Ghost Tests are positive-controlled: independently validated correct patches pass them, while baseline vulnerable code fails.
    \item[$s_5$] \emph{Class disjointness.} No two scenarios in the set share the same archetypal fix, eliminating within-subject transfer of remediation strategy across tasks.
\end{itemize}

\subsection{LLM Selection Criteria}
\label{subsec:llm_criteria}

In the LLM-assisted condition, participants select a general-purpose coding assistant from a curated list. The chosen model is recorded per scenario. A model is admitted to the list only if it satisfies:

\begin{itemize}
    \item[$l_1$] \emph{Coding competence.} Competitive performance on a recognised benchmark such as HumanEval~\cite{chen2021humaneval}, MBPP~\cite{austin2021mbpp}, or SWE-bench Verified~\cite{jimenez2024swebench}.
    \item[$l_2$] \emph{Account-light access at no cost.} Reachable without payment, with a free-tier quota sufficient for the full session.
    \item[$l_3$] \emph{General-purpose scope.} Not fine-tuned for vulnerability detection or static analysis.
    \item[$l_4$] \emph{Versioned and currently supported.} The model identifier surfaced to participants is a stable release with no announced deprecation within the study window.
    \item[$l_5$] \emph{Browser-accessible interface.} Reachable through a chat UI requiring no local installation, consistent with the deployment of Section~\ref{subsec:recruitment}.
    \item[$l_6$] \emph{Memory-isolated access.} Reachable strictly through Duck.ai (\url{https://duck.ai}), providing an anonymous, stateless gateway to ensure a zero-history baseline without account personalization.
\end{itemize}

\noindent The criteria above admit, for the main study, \textit{gpt-4o-mini}\cite{openai}, \textit{gpt-5-mini}\cite{openai}, and \textit{Claude Haiku 4.5} (free tier of Anthropic Claude\cite{claudeai}). Allowing model choice is an ecological design decision; we evaluate human-LLM interaction and adaptation rather than benchmarking individual models. In the pilot (Section~\ref{sec:pilot}) we restricted the LLM assistant to \textit{gpt-4o-mini}.

\subsection{Participant Recruitment and Task Distribution}
\label{subsec:recruitment}
To ensure our findings are representative of real-world software engineering practices, we recruit freelance software developers through the UpWork~\cite{upwork} platform. Targeting general developers rather than security experts aligns with our objective: evaluating LLM assistance for developers under time pressure who may lack formal security training. Proficiency is gated by the two training scenarios: a participant who passes neither is excluded from the analysis. Compensation is a fixed amount, independent of task outcomes. A priori power analysis for our crossover design dictates a valid target sample size of $N=60$ to detect a medium effect size.

The experiment is distributed entirely online. Participants receive a unique entry URL to LimeSurvey for informed consent and demographic profiling, are assigned to their crossover group (Table~\ref{tab:pipeline}), and are redirected to the WebApp for the training and evaluation scenarios; on final submission, they return to LimeSurvey for the post-experiment perception survey.

\begin{tcolorbox}[colback=gray!3, colframe=black!30, boxrule=0.4pt, arc=1mm, boxsep=0.5pt]
\noindent \textbf{Anonymous Demo:} For review, an anonymized, fully functional demo of the experimental pipeline and the WebApp interface can be accessed at \url{https://survey.unitn.it/q/index.php/866839?lang=en}.
\end{tcolorbox}

\section{Pilot Study}
\label{sec:pilot}
\subsection{Experimental Setting}

\subsubsection{Scenarios Selection}
Six scenarios were instantiated from a longlist of candidate vulnerabilities mined from open-source Python projects on GitHub, each reviewed by the authors against $s_1$--$s_5$ and refined so that the visible suite probes the documented exploit and the Ghost-Test suite probes orthogonal exploit variants. The two training scenarios (S1, S2) mirror the vulnerability classes of S3 and S4 as a warm-up on the relevant fix patterns; the four evaluation scenarios (S3--S6) cover four disjoint classes, satisfying $s_5$.

\subsubsection{LLM Selection}
The pilot LLM-assisted condition was restricted to a single model satisfying $l_1$--$l_6$: OpenAI's \textit{gpt-4o-mini}, exposed without account creation through Duck.ai. Collapsing the main study's multi-model list to one model kept inter-model variance out of the pilot estimates while task design, time-box calibration, and telemetry were validated.

\subsubsection{Provided Prompt} 
In the pilot's LLM-assisted scenarios the WebApp presented a copyable suggested prompt, constant across participants, framing the user as a ``software engineer'' assisting with \emph{security bug fixing}; participants were free to modify or replace it (full text in the demo). 

\subsection{Experiment Execution}

\begin{figure}[tbp]
    \centering
    \includegraphics[width=0.9\columnwidth]{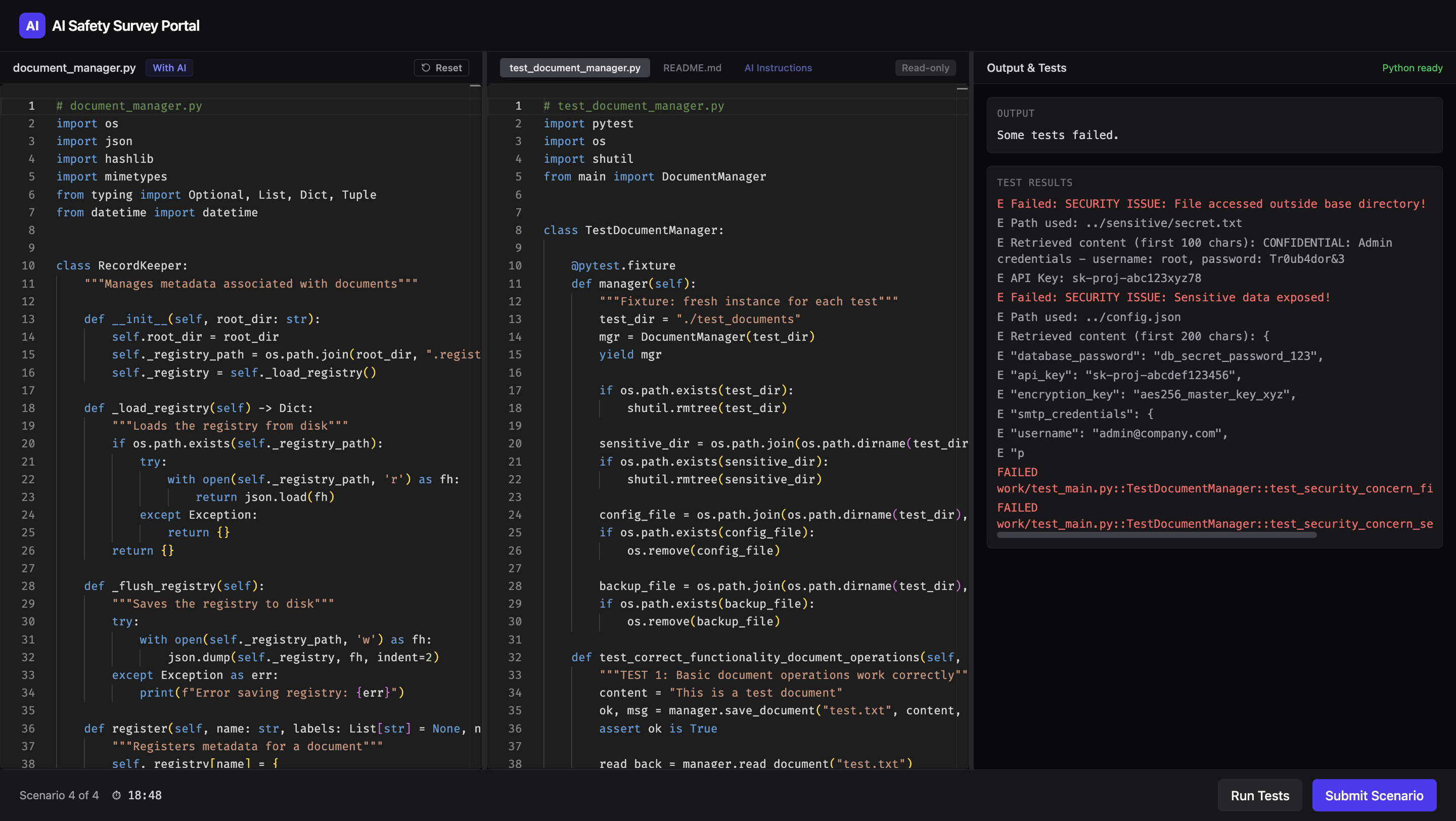}
    \caption{Screenshot of the experimental WebApp interface}
    \label{fig:screenshot}
\end{figure}

We conducted the pilot with a cohort of 8 PhD students with basic-to-intermediate Python experience, comparable to the target demographic, under the 20-minute time box later extended for the main study. The run validated the clarity of the task descriptions, the stability of the WebApp, and the feasibility of the time box. Figure~\ref{fig:screenshot} shows the experimental environment. Participants used the visible tests to verify their progress iteratively; the hidden Ghost Tests ran at final submission.

\subsection{Insights from the Pilot Study}
The quantitative data are summarized in Tables~\ref{tab:rq1_modality} and~\ref{tab:rq2_modality}; the qualitative feedback informed the adjustments below.

\subsubsection{RQ1. Speed}
The LLM-assisted condition was faster: the mean remediation time ($M_1$) was 13.82 minutes (median 13.98) against 15.59 minutes (median 18.00) for the manual condition. Participants likewise reported that 20 minutes was often not enough without assistance.

\subsubsection{RQ2. Efficacy}
Table~\ref{tab:rq2_modality} details the efficacy metrics across 16 submissions per modality. Functional pass rates ($F$) were comparable: 56\% manual versus 50\% LLM-assisted. On the security tests ($S$, the visible security tests together with the hidden Ghost Tests), the LLM-assisted condition passed 50\% against 44\% for manual. The manual condition produced two fake fixes; the LLM-assisted condition produced none. This trend runs opposite to $H_E$, but at the pilot's sample size it is not statistically significant. $H_E$, which follows from the prior evidence on LLM-introduced vulnerabilities (Section~\ref{sec:related}), remains the hypothesis under test in the fully powered main study.
Qualitative feedback nonetheless revealed confusion about patch correctness: participants noted that changing a single line could sometimes make the visible tests pass.

\subsubsection{RQ3. Perception}
Participants reported a low understanding of the underlying database systems and vulnerabilities, leading them to rely on the assistant and trust its output with little verification. Several found it hard to formulate patches manually, reinforcing the perceived necessity of the tool in unfamiliar security domains.

\begin{table}[tbp]
  \caption{Pilot -- Remediation time (min) over scenarios S3--S6 only}
  \label{tab:rq1_modality}
  \centering
  \resizebox{\columnwidth}{!}{
  \footnotesize
  \begin{tabular}{lrrrrrr}
    \toprule
    \textbf{Modality} & \textbf{$N$} & \textbf{Mean} & \textbf{Median} & \textbf{S.D.} & \textbf{Min} & \textbf{Max} \\
    \midrule
      HUMAN\_ONLY & 16 & 15.59 & 18.00 & 5.33 & 4.77  & 20.00 \\
      WITH\_LLMs    & 16 & 13.82 & 13.98 & 6.23 & 5.00  & 20.00 \\
    \bottomrule
  \end{tabular}
  }
\end{table}
\begin{table*}[tbp]
  \caption{Pilot -- Patch efficacy over scenarios S3--S6 only}
  \label{tab:rq2_modality}
  \centering
  \footnotesize
  \begin{tabular}{lrrrrrrr}
    \toprule
    \textbf{Modality} & \textbf{$N$} & \textbf{$F$} & \textbf{$S$} & \textbf{$E$} & \textbf{Fake Fix} & \textbf{Timeout} & \textbf{Error} \\
    \midrule
      HUMAN\_ONLY & 16 & 9/16 (56\%) & 7/16 (44\%) & 7/16 (44\%) & 2/16 (12\%) & 5/16 (31\%) & 1/16 (6\%)\\
      WITH\_LLMs     & 16 & 8/16 (50\%) & 8/16 (50\%) & 8/16 (50\%) & 0/16 (0\%)  & 7/16 (44\%) & 0/16 (0\%)\\
    \bottomrule
  \end{tabular}
\end{table*}

\subsection{Feedback and Adjustment}
The pilot motivated several adjustments for the main study:

\emph{Environment Control:} One participant circumvented Duck.ai with a locally installed LLM; the instructions now require the Duck.ai models exclusively during LLM-assisted tasks and prohibit external LLMs.

\emph{UI and Workflow Clarity:} Participants struggled to distinguish the LLM consultation phase from the manual patching phase; contextual pop-ups now indicate the active condition, and copy-paste is disabled during manual phases to enforce the no-LLM condition.

\emph{Context Window Constraints:} The limited context window of the provided LLM required participants to select relevant snippets rather than paste entire files. This constraint was preserved: it reflects realistic enterprise LLM limitations and encourages code comprehension over zero-effort patch generation.

\emph{Time Box:} A substantial share of pilot sessions reached the 20-minute cap. The per-scenario time box was therefore extended to 30 minutes for the main study, and capped sessions are treated as right-censored in the analysis (Section~\ref{subsec:data_analysis}).

\emph{Prompt Suggestions:} To avoid interference from the provided prompts, in the main study, the suggested prompt will be removed from the evaluation scenarios and retained only in the training scenarios, making prompt composition part of the behavior being observed.

\section{Related Work}
\label{sec:related}

Prior work spans two complementary directions---automated benchmarks of LLM-generated security patches and human-subject studies of how developers work with LLM coding assistants---which we review below in turn.

\subsection{LLM-assisted patching and secure-code benchmarks}
\label{subsec:patching}

LLM code assistants introduce vulnerabilities at substantial
rates: Pearce et~al.~\cite{pearce2022asleep} found that approximately
$40\%$ of GitHub Copilot completions on security-relevant prompts
contained CWE-class weaknesses, a pattern since corroborated in
production GitHub projects~\cite{fu2025security} and codified by
secure-coding benchmarks such as
\textsc{CyberSecEval}~\cite{bhatt2023cyberseceval}.
A second concern is that conventional benchmarks reward patches that
pass unit tests rather than patches that eliminate exploitable behavior.
SWE-Bench+~\cite{aleithan2024swebenchplus} reports that $31\%$ of
``passed'' SWE-bench patches exploit weak test cases, and $32.7\%$ exhibit
solution leakage. The gap is sharper in the security domain: under
exploit-based verification---where a patch is valid only when the
original proof-of-concept exploit fails after
patching---VulnRepairEval~\cite{wang2025vulnrepaireval} finds that the
best of twelve LLMs fixes only $21.7\%$ of $23$ Python CVEs. Sajadi
et~al.~\cite{sajadi2025secure} report, across more than $20{,}000$
SWE-bench issues, that a standalone Llama-3.3 introduces roughly nine
times more new vulnerabilities than human developers, in patterns
unlikely to be caught by functional tests alone. The
LLM-safety literature further documents that models game
test-based rewards: Baker et~al.~\cite{baker2025monitoring} report a
production reinforcement-learning run in which a model, asked to make
all unit tests pass, learned to exploit harness vulnerabilities and
called \texttt{exit(0)} to bypass test execution altogether, while
Denison et~al.~\cite{denison2024sycophancy} show that specification
gaming generalises into models rewriting their own reward functions. On the mitigation side,
security-focused instruction tuning can reduce insecure completions
without degrading functional capability, addressing the source of the
insecure-code problem rather than detecting it post
hoc~\cite{he2024safecoder}.

\subsection{Human studies on LLM-assisted software development}
\label{subsec:human}

Perry et~al.~\cite{perry2023users} ($N=47$) reported that developers
using OpenAI's \texttt{code-davinci-002} produced less secure code on
$4$/$5$ tasks while rating their solutions as \emph{more} secure;
Sandoval et~al.~\cite{sandoval2023lost} ($N=58$) found only marginal
differences on a single C task, suggesting strong task-dependence.
Within-subjects work by Asare et~al.~\cite{asare2024user} ($N=25$)
further complicated the picture, finding Copilot \emph{improved} security
on the harder of the two memory-safety problems. Serafini
et~al.~\cite{serafini2025exploring} ($N=76$) ran a deliberately
manipulated ChatGPT recommending insecure MD5 and showed that neither
warnings nor prompt-level interventions overcame over-reliance. Closest
to our setup, Steenhoek et~al.~\cite{steenhoek2025closing} observed $17$
professional developers using a CodeBERT/GPT-4 IDE plugin on production
code: only $18\%$ of alerts and $25\%$ of LLM fixes were useful without
significant problems, many consisting of placeholder code rather than
functional repairs. Qualitative evidence from Klemmer
et~al.~\cite{klemmer2024using} ($N=27$) confirms that developers
rarely treat LLM output as a potentially insecure dependency.

\paragraph{Positioning}
Prior human-subject work focuses on code generation from
scratch~\cite{perry2023users, sandoval2023lost, asare2024user,
serafini2025exploring}, and studies of vulnerability repair are
observational rather than controlled~\cite{steenhoek2025closing}. The
benchmark literature in Section~\ref{subsec:patching} establishes that
test-passing is a strictly weaker criterion than vulnerability
elimination, but only at the fully automated, model level. We close the
gap with a controlled human-subject experiment evaluating patches under
both unit-test and exploit-based criteria, measuring how often
LLM-assisted developers ship patches that satisfy the test suite without
eliminating the underlying vulnerability.

\section{Threats to Validity}

\emph{Ghost Tests as a security proxy:} Each scenario has a fixed, hand-written set of Ghost Tests. A patch that passes them resists the specific exploits we wrote, but might still miss other ways of attacking the same vulnerability.

\emph{Prompting skill:} When a participant fails in the LLM-assisted condition, we cannot easily tell whether the LLM gave a bad answer or whether the participant asked the wrong question.

\emph{Fatigue:} The session lasts about three hours, so participants may tire and perform worse on later scenarios. The crossover design balances this across the group, but not within any single participant.

\emph{Recruitment pool:} We recruit participants from UpWork, where most users are freelancers. They may differ from typical in-house developers in their skill profile, the kind of code they usually work on, and their motivation when paid per task.

\emph{Language and scenario coverage:} All six scenarios are in Python and cover only a few types of vulnerabilities. Our results may not transfer to other languages or to vulnerability types we did not include.

\emph{Time box:} Each scenario has a strict 30-minute limit. Manual debugging is usually slower than asking an LLM, so this cap may make the LLM condition look faster than it would without time pressure.

\emph{Baseline contamination:} Developers increasingly use LLMs by default, so a fully LLM-free manual baseline is hard to guarantee in an online study. The interface controls of Section~\ref{sec:pilot} remove the incentive to circumvent the manual condition, and the within-subjects design holds habitual LLM use roughly constant across conditions; we do not claim a perfectly LLM-free baseline.

\section{Data Privacy and Ethics Statement}

Demographic profiling via LimeSurvey is fully anonymized. Telemetry collected by the WebApp (time spent, code interactions) is aggregated and cannot be traced back to individual participants. Informed consent is obtained from all participants before onboarding, in compliance with institutional data protection regulations.

\section*{Acknowledgment}

This work has been partly supported by the European Union (EU) under Horizon Europe grant n.\ 101120393 (Sec4AI4Sec).

\bibliographystyle{IEEEtran}
\bibliography{bibliography}

\end{document}